\documentclass[amssymb,amsmath,superscriptaddress,footinbib,twocolumn]{revtex4}
\usepackage{natbib,graphicx,subfigure,subeqnarray,fancyhdr,amsmath,multirow,color, mathrsfs}
\begin{document}
\def\F{{\bf F}}\def\Re{{\rm Re}}
\def\btau{\boldsymbol{\tau}}
\def\bom{\boldsymbol{\omega}}
\def\bOm{\boldsymbol{\Omega}}
\def\bsigma{\boldsymbol{\sigma}}
\def\bSigma{\boldsymbol{\Sigma}}
\def\gamd{\dot{\boldsymbol{\gamma}}}
\def\u{{\bf u}}\def\n{{\bf n}}
\def\a{{\bf a}}\def\f{{\bf f}}\def\G{{\bf G}}
\def\A{{\bf A}}\def\B{{\bf B}}\def\C{{\bf C}}\def\D{{\bf D}}
\def\v{\vspace {2cm}}
\def\t{{\bf t}}\def\q{{\bf q}}\def\j{{\bf j}}
\def\U{{\bf U}}\def\V{{\bf V}}
\def\R{{\bf R}}\def\L{{\bf L}}\def\Real{{\cal R}}
\def\E{{\bf E}}\def\X{{\bf X}}
\def\bepsilon{\boldsymbol{\epsilon}}
\def\r{{\bf r}}\def\x{{\bf x}}\def\d{{\rm d}}\def\dd{{\bf d}}
\def\e{{\bf e}}\def\1{{\bf 1}}\def\0{{\bf 0}}
\def\p{{\partial}}\def\pp{{\bf p}}\def\k{{\bf k}}
\def\De{{\rm De}}\def\Sp{{\rm Sp}}
\def\h{\noindent\rule{\textwidth}{0.4pt}}
\def\S{{\bf S}}
\def\dgamd{\stackrel{\triangledown}{\gamd}}
\def\dbtau{\stackrel{\triangledown}{\btau}}
\newcommand{\icomp}{\mbox{i}}
\newcommand{\icomps}{\mbox{\scriptsize i}}
\newcommand{\expo}{\mbox{e}}
\newcommand{\eps}{\epsilon}
\newcommand{\veps}{\varepsilon}
\newcommand{\vecI}{\mbox{\boldmath$I$}}
\newcommand{\sst}[1]{\scriptscriptstyle{#1}}
\newcommand{\vecu}{\mbox{\boldmath$u$}}
\newcommand{\vecp}{\mbox{\boldmath$p$}}
\newcommand{\vecr}{\mbox{\boldmath$r$}}
\newcommand{\vecS}{\mbox{\boldmath$S$}}
\newcommand{\vecG}{\mbox{\boldmath$\mathcal{G}$}}
\newcommand{\Nab}{\mbox{\boldmath$\nabla$}}
\newcommand{\perps}{{\mbox{$\scriptscriptstyle{\perps}$}}}
\newcommand{\bk}{\bar{k}_x}
\newcommand{\bkD}{k_x\Delta}

\title{Clustering instability of focused swimmers}
\author{Eric Lauga}
\email{e.lauga@damtp.cam.ac.uk}
\affiliation{Department of Applied Mathematics and Theoretical Physics, 
University of Cambridge,  Cambridge CB3 0WA, United Kingdom.}
\author{Francois Nadal}
\email{f.r.nadal@lboro.ac.uk}
\affiliation{Department of Mechanical, Electrical and Manufacturing Engineering, Loughborough University,Loughborough LE11 3TU, United Kingdom}
\date{\today}

\begin{abstract}
One of the hallmarks of active matter is its rich nonlinear dynamics and instabilities. Recent numerical simulations of phototactic algae 
showed that a thin jet of swimmers, obtained from hydrodynamic focusing inside a Poiseuille flow, was unstable to longitudinal perturbations 
with swimmers dynamically clustering (Jibuti et al., Phys. Rev. E, {\bf 90}, 2014). As a simple starting point to understand these instabilities, we consider in this paper an initially homogeneous one-dimensional line of aligned swimmers moving along the same direction, and characterise its instability using both a  continuum framework and a discrete approach. In both cases, we show that hydrodynamic interactions between the swimmers lead to instabilities {in density} for which we compute the growth rate analytically. Lines of pusher-type swimmers are predicted to remain stable while lines of 
pullers (such as flagellated algae) are predicted to always be unstable.
\end{abstract}
\maketitle

\section{\bf Introduction}

A  fascinating recent development in soft condensed matter physics  is the flurry of new results on   active matter \cite{ramaswamy10}.
Originally motivated by the over-damped limit of  swimming microoganisms \cite{lp09}, the physics of active matter also encompasses  
active gels,  driven granular suspensions, filament-motor protein complexes and the cytoskeleton of eukaryotic cells \cite{marchetti13}.

One of the important issues in  active matter research  is  that  of pattern formation and instabilities. Under which conditions does a 
particular homogeneous, isotropic  system remain stable and what parameters govern its transition to a fluctuating, inhomogeneous state? 

The question of stability has been the subject of many theory papers in the case of swimming cell suspensions. Aligned three-dimensional 
suspensions of swimmers are always unstable  to density and orientation perturbations \cite{simha02,saintillan07,saintillan08}. In 
contrast,  homogeneous, isotropic suspensions  are linearly unstable  to long wavelengths perturbations in orientation for pushers-type  
cells (swimming cells propelled from their back, such as flagellated bacteria) but stable for pullers-type cells (cells propelled from 
their front, such as flagellated algae) \cite{saintillan08,hohenegger10,koch}. 

Beyond  stability, many studies have looked to characterise the  nonlinear, collective dynamics of swimming cells, both computationally 
\cite{hernandez-ortiz05} and  experimentally \cite{sokolov12}, and have shown how collective modes of locomotion could lead to enhanced 
transport in the surrounding fluid  \cite{wu00,valeriani11,jepson13,kasyap14} and mixing \cite{pushkin13prl}, novel rheological 
characteristics \cite{chen07,sokolov09} and could power synthetic systems \cite{sokolov10}.

\begin{figure}[t]
\begin{center}
\includegraphics[width=.39\textwidth]{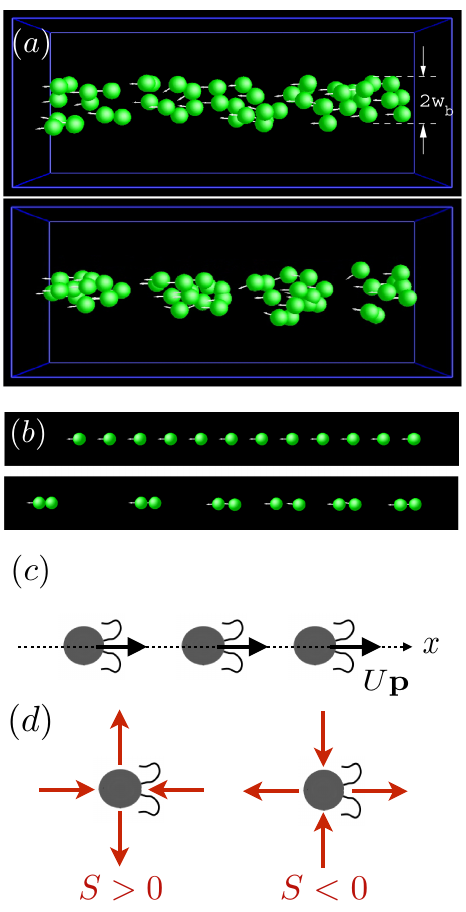}
\caption{(a) Clustering instability of phototactic algae from numerical simulations 
as main motivation for our work \cite{jibuti14}; (b) Similar instability arising on a one-dimensional line of swimming algae,  which is the precise setup considered in this paper.   
Reproduced  from Jibuti et al.~(2014) Phys. Rev. E, {\bf 90}. Copyright 2014, American Physical Society; (c) Sketch of the model problem addressed in this paper: a one-dimensional line of swimmers of fixed orientation  $\pp=\e_x$  swimming with identical speed $U$ in the absence of rotational diffusion; (d) each swimmer acts on the fluid as a stresslet of magnitude $S$, creating in the swimming frame local flows as illustrated by  solid red arrows.
}
\label{fig1}
\end{center}
\end{figure}

Recently, a numerical study addressed the dynamics of  a suspension of  phototactic algae i.e.~cells whose direction of motion was set 
by the presence of an external light source. When present in a pressure-driven (Poiseuille) flow, these swimming cells hydrodynamically 
focus into a thin jet in the center of the channel (itself a classical result \cite{pedley92}) which was shown to be unstable to  
longitudinal perturbations  with swimmers clustering along the jet \cite{jibuti14}. This instability is illustrated in Fig.~\ref{fig1}a 
and a similar instability was observed numerically  in the case of one-dimensional lines of model algae (Fig.~\ref{fig1}b). 

{As a simple starting point to understand these instabilities, we consider in this paper the  problem of an initially homogeneous one-dimensional line of aligned swimmers moving along the same direction. We ignore rotational diffusion (except in the introduction section where the standard theoretical framework is summarised) and therefore the swimmers remain aligned, while their one-dimensional density is allowed to vary. We  characterise analytically density instabilities  using  two complementary 
modelling approaches, namely a continuum framework and a discrete, point-swimmer, framework.} In both cases we show that 
hydrodynamic interactions between the swimmers are responsible for the clustering and compute the growth rate of the instability. 
Both approaches give the same results and indicate that the instability arises only for puller-type swimmers such as the algae 
considered in Ref.~\cite{jibuti14} while pushers are predicted to remain stable.

\medskip
\section{\bf Continuum framework}

In the first approach, we  use the classical continuum framework for suspension of swimming cells \cite{simha02,saintillan13}. 
The suspension is characterised by a probability  distribution function, $\Psi(\r,\pp,t)$, for the position $\r$ and orientation 
$\pp$  of the cells. Conservation of probability is written as
\begin{equation}
 \frac{\p \Psi}{\p t}+ \nabla_\r\cdot(\dot\r \Psi )+ \nabla_\pp\cdot(\dot\pp \Psi ) =  D \nabla_\r^2 \Psi + 
  D_R \nabla_\pp^2 \Psi,
\end{equation}
where $D$ is a diffusion constant in position and $D_R$ in orientation.  The swimmers self-propel   with speed $U$ along their 
direction $\pp$ {(Fig.~\ref{fig1}c)} and are also advected by the flow $\u$ and thus we have the relationship
\begin{equation}
\dot \r = U \pp  + \u.
\end{equation}
The velocity field $\u$ is incompressible, $
 \nabla_\r \cdot \u = 0$, and  satisfies Stokes equation with a pressure field $q$
\begin{equation}
-\nabla_\r q + \mu \nabla_\r ^2 \u +   \nabla_\r\cdot \bSigma 
 = \0,
\end{equation}
the last term being the active stress due to the active particles, namely the stresslet
\begin{equation}
\bSigma=S \left\langle\pp \pp-\frac{1}{3}{\bf 1} \right\rangle,
\label{eq:active_stress}
\end{equation}
{where the brackets $\langle \cdot \rangle$ denote  orientational average i.e.
\begin{equation}
\langle \cdot \rangle \equiv \int \Psi(.) \d\pp.
\end{equation}}
In Eq.~(\ref{eq:active_stress}), the  sign of the stresslet  strength, $S$, plays an important role in collective dynamics: 
$S<0$ for pushers and $S>0$ for pullers {(see illustration in Fig.~\ref{fig1}d).} 

In the simulation of Ref.~\cite{jibuti14}, the swimmers are phototactic and their swimming orientation is identical and aligned 
with the line of swimmers.  We thus assume here that the swimmers are located on a single line, along the $x$ direction, and that 
their orientation is fixed and also along $x$. 
We thus look for solutions of the form
\begin{equation}\label{Psi_line}
\Psi = \delta (y)\delta(z)\lambda (x,t) \delta (\pp - \e_x),
\end{equation}
where $\lambda$ is the one-dimensional density of swimmers along the $x$ axis. 
{Note that for the particular solution of the form of Eq.~\eqref{Psi_line} to be admissible, we ignore diffusion in orientation in what follows as well as diffusion in position in the directions perpendicular to the axis $x$.} 
Given the fixed orientation, the active stress is given by 
\begin{equation}
\bSigma = S\int\Psi \left(\e_x\e_x - \frac{1}{3}{\bf 1}\right) \d \pp,
\end{equation}
which, using Eq.~\eqref{Psi_line}, becomes 
\begin{equation}
\bSigma = S \delta (y)\delta(z)\lambda (x,t) \left(\e_x\e_x - \frac{1}{3}{\bf 1} \right).
\end{equation}

The flow equation becomes
\begin{eqnarray}
\frac{\p q}{\p x} &=& \mu \nabla^2 u_x + \frac{2}{3}S\frac{\p }{\p x}\left[  \delta (y)\delta(z)\lambda (x,t)\right],\label{debut}\\
\frac{\p q}{\p y} &=& \mu \nabla^2 u_y -\frac{1}{3}S\frac{\p }{\p y}\left[  \delta (y)\delta(z)\lambda (x,t)\right],\\
\frac{\p q}{\p z} &=& \mu \nabla^2 u_z - \frac{1}{3}S\frac{\p }{\p z}\left[  \delta (y)\delta(z)\lambda (x,t)\right],\\
0 & = & \frac{\p u_x}{\p x}+\frac{\p u_y}{\p y}+\frac{\p u_z}{\p z}\cdot\label{fin}
\end{eqnarray}

In order to make progress, we use  Fourier transforms defined for any function $\alpha (\x,t)$ of space and time as
\begin{equation}
\alpha (\x,t) =\frac{1}{(2\pi)^3}\int e^{i(\k\cdot\x)}\hat\alpha (\k,t)\d \k. 
\end{equation}
We proceed by  Fourier-transforming Eq.~\eqref{debut}-\eqref{fin} in space leading to
\begin{eqnarray}
ik_x \hat q &=& -\mu k^2 \hat u_x + \frac{2}{3}S ik_x \hat \lambda,\label{debut2}\\
ik_y \hat q &=& -\mu k^2 \hat u_y -\frac{1}{3}Sik_y \hat\lambda,\\
ik_z \hat q &=& -\mu k^2 \hat u_z - \frac{1}{3}Sik_z \hat\lambda,\\
0 & = &\k \cdot \hat \u.
\label{fin2}
\end{eqnarray}

In order to determine the pressure we take the dot product of the $\k$ vector with the first three equations and exploit Eq.~\eqref{fin2} to get
\begin{equation}
ik^2 \hat q = S i \hat \lambda \left[\frac{2}{3}k_x^2 - \frac{1}{3}(k_y^2 + k_z^2)\right] = Si\hat\lambda 
\left[k_x^2 - \frac{1}{3}k^2\right].
\label{q}
\end{equation}
Plugging Eq.~\eqref{q} into Eq.~\eqref{debut2} then leads to 
\begin{equation}
i\frac{k_x}{k^2} 
S\hat\lambda 
\left[k_x^2 - \frac{1}{3}k^2\right]
=-\mu k^2 \hat u_x + \frac{2}{3}S ik_x \hat \lambda,
\end{equation}
which can be solved for $\hat u_x$ as
\begin{equation}
\hat u_x = \frac{iS\hat\lambda}{\mu} \frac{k_x  }{ k^2 }\left(
1 -\frac{k_x^2}{k^2}  \right).\label{ux_hat}
\end{equation}

In order to close the stability calculation we finally need to write down conservation of swimmers along the line. In one dimension 
the conservation of swimmers is written as 
\begin{equation}
\frac{\partial \lambda}{\partial t} + \frac{\partial }{\partial x}\left[(U  + u_x|_{(x,0,0,t)}  )\lambda \right] = D 
\frac{\partial^2 \lambda}{\partial x^2} \cdot \label{conservation}
\end{equation}

Note that using Fourier notation, we can see that
\begin{equation}\label{u_0}
u_x|_{(x,0,0,t)} =\frac{1}{(2\pi)^3}\int e^{ik_x x}\hat u_x (\k,t) \d \k.
\end{equation}

The base case is a uniform distribution of swimmers, characterised by $\lambda (x,t) =\lambda_0$, so that 
$\hat \lambda = \lambda_0 \delta(k_x)$. {Note that $\lambda_0$ is the inverse of the initial inter-swimmer distance, which we denote
$\Delta$ in what follows}. From Eq.~\eqref{u_0} and using Eq.~\eqref{ux_hat} we thus see that  
\begin{equation}
u_x|_{(x,0,0,t)} = \int e^{ik_x x} 
 \frac{iS \lambda_0 \delta(k_x)}{\mu} \left(
\frac{k_x  }{ k^2 } -\frac{k_x^3}{k^4}  \right)
\d \k=0,
\end{equation}
and therefore the uniform concentration is indeed a base state. 

We then  consider perturbations of this case state, written as $\lambda (x,t) =\lambda_0 + \lambda'$, $u_x = u_x'$. The linearisation of 
Eq.~\eqref{conservation} around the base state gives 
\begin{equation}
\frac{\partial \lambda'}{\partial t} 
+U \frac{\partial \lambda'}{\partial x}   
+\lambda _0 \frac{\partial }{\partial x}\left[u_x'|_{(x,0,0,t)}   \right] 
= D \frac{\partial^2 \lambda'}{\partial x^2} \cdot \label{conservation'}
\end{equation}

To get a self contained equation for $\lambda'$ we then use the definition of the Fourier integral and rewrite it as
\begin{equation}
\frac{\partial \lambda'}{\partial t} 
+U \frac{\partial \lambda'}{\partial x}  
+ i \frac{\lambda _0 k_x}{(2\pi)^3} 
\int e^{ik_x x}\hat u'_x (\k,t) \d \k
= D \frac{\partial^2 \lambda'}{\partial x^2} \cdot 
\end{equation}
Fourier transforming along the $x$ direction gives
\begin{equation}
\frac{\partial \hat \lambda'}{\partial t} 
+U i k_x  \hat \lambda'
+ \frac{i \lambda _0 k_x }{(2\pi)^{2}}
\int \hat u'_x (\k,t) \d k_y\d k_z
= - D k_x^2 \hat  \lambda'  .
\end{equation}
and then using Eq.~\eqref{ux_hat} leads to the final stability relationship
\begin{equation}
\frac{\partial \hat \lambda'}{\partial t} 
+U i k_x  \hat \lambda'
= \frac{S k_x^2     \lambda _0 \lambda'}{\mu(2\pi)^{2}}
\int 
\left(
1 -\frac{k_x^2}{k^2}  \right)
 \frac{ \d k_y\d k_z}{ k^2 }
- D k_x^2 \hat  \lambda'  
. 
\end{equation}
Looking for
exponentially growing modes of the form $\hat\lambda ' (k_x,t)=f(k_x)e^{\sigma t} $ and we find a dispersion relation 
\begin{equation}\label{disp:cont}
\sigma =-  iUk_x
+ \frac{S k_x^2     \lambda _0 }{\mu (2\pi)^{2}}
\int 
\left(
1 -\frac{k_x^2}{k^2}  \right)
 \frac{ \d k_y\d k_z}{ k^2 }
- D k_x^2 . 
\end{equation}
This dispersion relationship has three terms. The first is a pure  traveling mode reflecting the fact that 
we are in the laboratory frame while the swimmers move with speed $U$. The third term  is diffusive and stabilizing. 
In contrast, the second term is the one arising from hydrodynamic interactions and is the one leading to instabilities. 

{We can evaluate the integral in the second term of Eq.~\eqref{disp:cont} using cylindrical coordinates,
writing $k_{\perp}^2 = k_y^2 + k_z^2$, which yields
\begin{align}
I = \int \frac{1}{ k^2 }\left(1 -\frac{k_x^2}{k^2}\right)&\d k_y\d k_z= 2\pi \int_{k_{\mbox{\tiny min}}}^{k_{\mbox{\tiny max}}}\!\!\!\! 
\frac{k_{\perp}^3}{(k_{\perp}^2+k_x^2)^2} \d k_{\perp},
\label{eq:disp_continuous_1}
\end{align}
where the bounds $k_{\mbox{\tiny min}}$ and $k_{\mbox{\tiny max}}$ of the second  integral must be specified. We assume for simplicity that the swimmers move in an unbounded space such that the lower bound $k_{\mbox{\tiny min}}$ can be taken equal to zero. The upper bound needs to scale with $\Delta^{-1}$ since below the  length scale $\Delta$ the continuous approach loses its meaning; for simplicity we take it here to be 
$k_{\mbox{\tiny max}} = \Delta^{-1}$ and making a different choice of the form $k_{\mbox{\tiny max}} = \alpha \Delta^{-1}$ does not affect the main  results below.

The integral in Eq.~(\ref{eq:disp_continuous_1}) can then be calculated exactly and one finds
\begin{equation}
\sigma = -  iUk_x + \frac{1}{4\pi}\frac{Sk_x^2}{\Delta\mu}\,f(k_x\Delta) - Dk_x^2,
\label{eq:disp_continuous_2}
\end{equation}
where the function $f$ is given by
\begin{equation}
f(u) \equiv -\left[\ln\left(\frac{u^2}{1+u^2}\right) + \frac{1}{1+u^2}\right].
\end{equation}

To capture the instability, we need to consider only the real part, $\sigma_R$, of the right-hand side of Eq.~(\ref{eq:disp_continuous_2}), i.e.~its last two terms. 
At small wavenumbers, i.e.~for wavelength  much larger than the initial (mean) distance between swimmers $\bkD \ll 1$, 
the real part of the growth rate scales as 
\begin{equation}
\sigma_R (k_x) \sim -\frac{1}{2\pi}\frac{S}{\mu\Delta^3}(\bkD)^2\ln(\bkD),
\label{eq:scaling_small_k_cont}
\end{equation}
and diffusion plays no role. A similar result will be obtained below with a discrete approach. 

As can be seen from Eq.~\eqref{eq:scaling_small_k_cont},  in the case of {puller cells} with $S>0$, since the $\log $ term is negative, the real part of the growth rate is positive, and long wavelengths are unconditionally unstable. As discussed below, when considering  practical situations of real microscopic swimmers, small wavelengths are also unstable, and thus a whole band of  wavelengths $[0,\Delta^{-1}]$ is unstable, a result  in line with Jibuti's simulations\cite{jibuti14}. Conversely, for {negative} values of $S$, so-called pusher cells, 
the growth rate is always negative and the system is always stable, also in agreement with Jibuti's results.

\section{\bf Discrete framework}

We now consider a second, complementary, modelling approach for the same problem. 

Specifically, we model the three-dimensional  
swimmers as discrete moving  stresslets  located on a uniform one-dimensional lattice of spacing $\Delta$.  {Due to the long range nature of the hydrodynamic interactions, we include the flow created by all  other swimmers when computing the velocity of  a given swimmer. The swimmers are equally spaced on the $x$ axis and we wish to assess  the stability of such a situation.  
In the reference frame moving with speed $U$ along the $x$ axis, the swimmers are motionless if the situation is  unperturbed.  We then subject the homogenous line to a harmonic perturbation of wavenumber
$k_x = 2\pi/(p\Delta)$, where $p$ a positive integer, such that each individual swimmer is shifted from its equilibrium position by a small quantity
$\epsilon_n \equiv \veps\,\cos(k_x\,n\Delta) = \veps\,\cos(2\pi n/p)$ (see notation in  Fig.~\ref{swimmers_chain}). Since the wavelength of the perturbation needs to be larger than $2\Delta$,
the wavevector is bounded by $\pi/\Delta$.
\begin{figure}[t!]
\includegraphics[width=.4\textwidth]{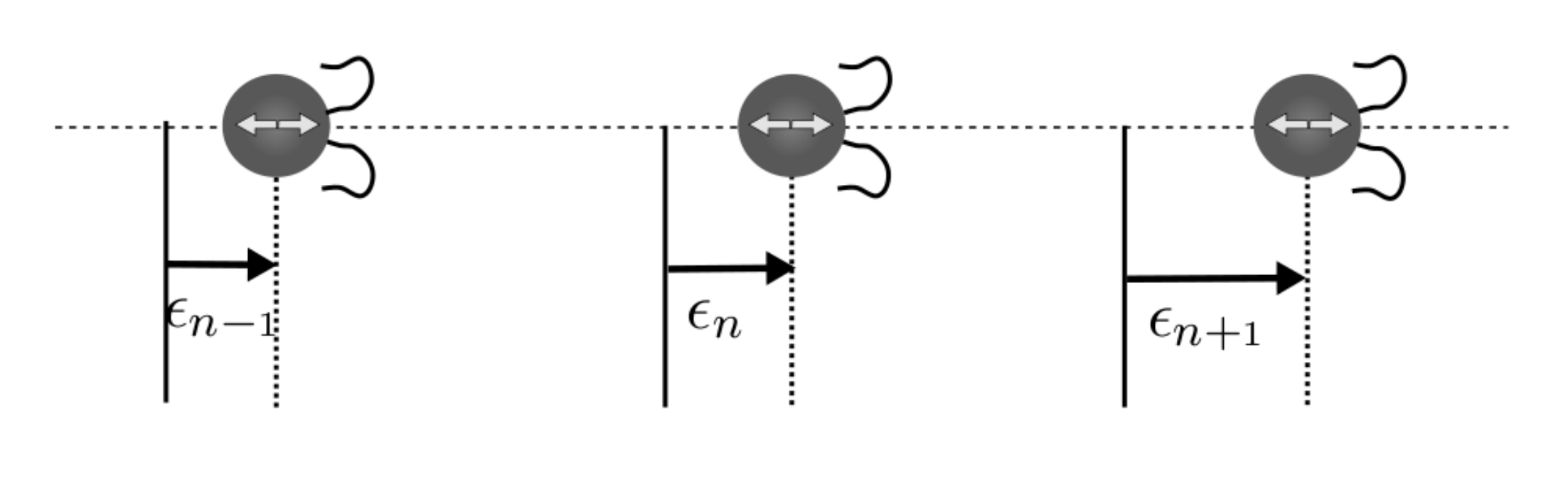}
\caption{Discrete homogeneous one-dimensional line of swimmers. In the  frame moving with the  swimming speed of the cells, the 
perturbation distance to the equilibrium steady position of each swimming, $x_n$,
is denoted  $\eps_n$.}
\label{swimmers_chain}
\end{figure}

A swimmer located at $\r_{\sst 0}$ generates its own flow given by a force dipole
\begin{equation}
\vecu^d (\r-\r_{\sst 0}) = \vecS(\pp)\colon\!\!\Nab\vecG(\r - \r_{\sst 0}),
\label{eq:dipole_flow}
\end{equation}
where $\vecS(\pp) = S\,\pp\pp$ and $\vecG(\r) = (\1 + \r\r)/8\pi\mu r$ is the stokeslet (point force) solution.
All the dipoles have the same fixed orientation and subject to the flow generated on the $x$-axis
($y = z = 0$) by all other swimmers. The $x$-component of the velocity field, Eq.~(\ref{eq:dipole_flow}), reduces to
\begin{equation}
u^d(x-x_{\sst 0}) = -\frac{S}{4\pi\mu}\frac{x-x_{\sst 0}}{|x-x_{\sst 0}|^3}\cdot
\label{eq:dipole_flow_1D}
\end{equation}

In the absence of inertia, the dynamics of swimmer \#$n$ is governed by the evolution of its perturbation, away from the
equilibrium position, $\epsilon_n$, which follows thus
\begin{equation}
\dot{\eps}_n = \sum_{q>1}\left[u^d(x_n - x_{n-q}) +  u^d(x_n - x_{n+q})\right],
\label{eq:dynamics_swimmer}
\end{equation}
where the infinite sum includes interactions with all swimmers. 
The first term on the right-hand side of Eq.~(\ref{eq:dynamics_swimmer}) can be expanded to first order in $\eps_n -\eps_{n-q}$  
since $x_n - x_{n-q} = \Delta + \eps_n - \eps_{n-q}$ with $(\eps_n - \eps_{n-q})/\Delta \ll 1$, and one gets at first order
\begin{equation}
u^d(x_n - x_{n-q}) = -\frac{S}{4\pi\mu}(q\Delta)^{-2}\left(1-2\frac{\eps_n-\eps_{n-q}}{q\Delta}\right),
\label{eq:expand_1}
\end{equation}
{with a similar result for $u^d(x_n - x_{n+q})$}. 
Introducing these expansions  in Eq.~(\ref{eq:dynamics_swimmer}) leads to the equation which governs
the evolution of the $\eps_n$ in time,
\begin{equation}
\dot{\eps}_n = -\frac{S}{4\pi\mu} \sum_{q>1}\frac{1}{(q\Delta)^3}(\eps_{n+q} + \eps_{n-q} - 2 \eps_n).
\label{eq:dynamics_swimmer_exp}
\end{equation}
Note that this equation is valid in both the laboratory frame or the frame moving with the swimming  speed $U$. 

The perturbation in displacement $\eps_n$ (which ``follows'' the global displacement of the assembly at speed $U$) can be considered 
as a propagative wave in the laboratory frame, so that we consider a discrete perturbation of the form
\begin{equation}
\eps_{n\pm q} = \veps\,\expo^{\tilde \sigma t}\expo^{-\icomps[k_x (n\pm q)\Delta-\omega t)]},\;\;\mbox{with}\;\omega = k_xU.
\end{equation}
When introduced in Eq.~(\ref{eq:dynamics_swimmer_exp}), this leads to the dispersion relationship which provides
the discrete growth rate $\tilde \sigma(k_x)$, and one finds the infinite sum
\begin{equation}
\tilde \sigma(k_x) = -\icomp U k_x + \frac{2}{\pi}\frac{S}{\mu\Delta^3}\,\sum_{q>1}\frac{\sin^2(q k_x\Delta/2)}{q^3}\cdot
\label{eq:disp_discrete}
\end{equation}
Here again, we see clearly from Eq.~\eqref{eq:disp_discrete} that if $S<0$ (pushers) the system is predicted to be stable while a line suspension of pullers ($S>0$) is always unstable.

To address the behaviour at long wavelengths, we rewrite the  real part $\tilde \sigma_{R}$ of the discrete growth rate, second term in the right-hand side of Eq.~(\ref{eq:disp_discrete}),  in the following form
\begin{equation}
\tilde \sigma_{R}(k_x) = \frac{1}{2\pi}\frac{S}{\mu\Delta^{3}} (\bkD)^2\,\sum_{q>1}\frac{\sin^2(q \bkD /2)}{(q\bkD/2)^3}\,\frac{\bkD}{2}.
\label{eq:disp_discrete_2}
\end{equation}
The sum in Eq.~(\ref{eq:disp_discrete_2}), which has the form of a discrete integral, is bounded as follow
\begin{align}
\int_{\bkD/2}^\infty\!\!\frac{\sin^2u}{u^3}\,du & < \notag\\
\sum_{q>1}&\frac{\sin^2(q \bkD /2)}{(q\bkD/2)^3}\,\frac{\bkD}{2}\notag\\
& < \mbox{sinc}^2(\bkD/2) + \int_{\bk/2}^\infty\!\!\frac{\sin^2u}{u^3}\,du.\label{eq:sum_bound}
\end{align}
Given that the integral in both right and left bounds of the previous double inequality scales as $-\ln \bkD$ for $\bkD\rightarrow 0$, one obtains the
scaling of $\sigma_{R}$ at small wavenumbers, namely
\begin{equation}
\tilde \sigma_{R}(k_x)\sim -\frac{1}{2\pi}\frac{S}{\mu\Delta^3}(\bkD)^2\ln(\bkD),
\label{eq:scaling_small_k_disc}
\end{equation}
which is exactly  the same relationship as the one obtained in  the continuum limit, Eq.~(\ref{eq:scaling_small_k_cont}).
}

.

\section{\bf Discussion}

The spatial organisation of active matter under the combined effects of external Poiseuille flows and physical taxis (such 
as magnetotaxis or phototaxis) has been the subject of recent numerical and experimental studies \cite{jibuti14,Waisbord2016,Martin2016}. 
The interplay between hydrodynamic  effects and physical taxis classically results in a focusing of the swimmers close to the axis 
of the  channel \cite{pedley92}, the radial density profile being determined by the competition between the swimming-enhanced
diffusivity of the swimmers and the amplitudes of external forcing.

In this paper, we examined theoretically the  linear stability of swimmers along a one-dimensional clusters from both a continuum and 
discrete perspectives. The continuum approach, performed in Fourier space, leads to a stability condition involving two competing 
terms coming from hydrodynamic interactions and diffusion, respectively.

At small $\bkD$, both modelling approaches provide a real part of the
growth rate scaling as $\sigma_R(k_x)\sim -(1/2\pi)[S/(\mu\Delta^3)](\bkD)^2\ln(\bkD)$. In this limit diffusion plays no role --  
a  result is in fact true, with very good approximation, for all wavenumbers.  To see this, we take $D/\Delta^2$ as a typical timescale and denoting $\bar{k}_x$ the dimensionless product $k_x\Delta$,    
the real part $\sigma_{R}$ of the growth rate obtained in the continuum limit can be rewritten in the following dimensionless form
\begin{equation}
\bar{\sigma}_R(\bar{k}_x) = \bk^2\left[\bar{S}\,f(\bk)-1\right],
\label{eq:growth_rate_real_adim}
\end{equation}
where $\bar{\sigma}_R = \Delta^2 \sigma_R/D$ and $\bar{S} = (1/4\pi)[S/(D\Delta\mu)]$.
For  microscopic swimmers of size $a = 1$~$\mu$m separated by an initial distance $\Delta/a = 2$ and propelled in water
($\mu = 10^{-3}$ Pa.s) by an individual stresslet of $S = 1$ pN.$\mu$m, and writing  $D\mu = k_{B}T/6\pi a$ where $k_{B}$ and $T$ refer to the Boltzmann constant and the temperature (300 K), one obtains the dimensionless value $\bar{S} \simeq 181$. Given the typical value of the dimensionless function $f$,  one obtains that the diffusive term in Eq.~\eqref{eq:growth_rate_real_adim} can basically always be neglected  and with a good approximation we  can write
\begin{equation}
\sigma_R(k_x) \simeq \frac{1}{4\pi}\frac{S}{\mu\Delta^3}\,(\bkD)^2\,f(\bkD).
\label{eq:simpl_growth_rate_real_adim}
\end{equation}

\begin{figure}
\begin{center}
\includegraphics[width=.36\textwidth]{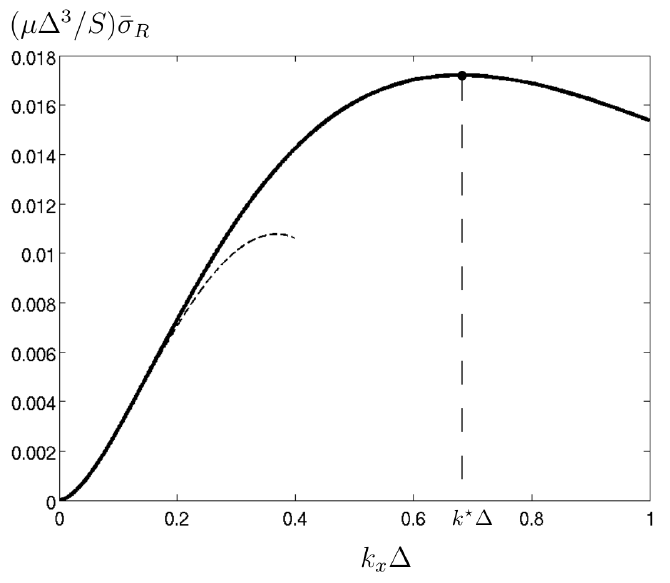}
\end{center}
\caption{Dimensionless real part of the growth rate as a function of the dimensionless wave number, in the large $\bar{S}$ limit,  Eq.~(\ref{eq:simpl_growth_rate_real_adim}). The dashed lines correspond to the 
improved small $\bkD$ scaling $\sigma_R(k_x) \simeq -(1/2\pi)[S/(\mu\Delta^3)](\bkD)^2\,[\ln(\bkD) + 1/2]$ obtained by  taking into account the constant in the Taylor expansion of the function $f$.
}
\label{fig:sigma_k_dim}
\end{figure}

The real part of the growth rate, non-dimensionalised  as  $S/(\mu\Delta^3)$,  is plotted 
in the large $\bar{S}$ limit (where diffusion is negligible) as a function of the dimensionless axial wavenumber, $k_x\Delta$, 
in Fig.~\ref{fig:sigma_k_dim}, for {positive} values of the  stresslet strength $S$
(i.e.~in the unstable case of pullers). All wavelengths $k_x\Delta \in [0,1]$ are seen to be  unstable. Furthermore, a most unstable wavelength $k^{\star}\Delta\approx 0.68$  is obtained. Considering however the small difference between the growth rate at $k^\star\Delta$ and at $k_x\Delta = 1$,
no real emergence of the most unstable wave length $\lambda^\star = 2\pi/k^\star$ should be expected experimentally. This is consistent with the
``pairing'' scenario depicted in Jibuti et al.~\cite{jibuti14}. It is likely that this pairing would continue in sequence, with  swimmer pairs, which also act at pullers, pairing up, eventually leading to one big cluster.

The physical mechanism leading to the instability captured in our paper is in fact quite elementary, and can be captured by considering the line of swimmers sketched in Fig.~\ref{swimmers_chain}. Puller cells induce attractive flows along their swimming axis with a magnitude which increases as one gets closer to the cell (Fig.~\ref{fig1}d). In contrast, pushers induce repulsive flows, also with a magnitude increasing near the cell (Fig.~\ref{fig1}d). Consider a situation where  the location of the middle cell in Fig.~\ref{swimmers_chain} is perturbed to its right. If the cells are pushers, the repulsion with its neighbour on the right increases while the repulsion with the cell on the left decreases, and the cell returns to its original location, indicating stability. If in contrast the cells are pullers, the attraction toward the cell on the right increases, and the attraction with the left on the left decreases, leading to an amplification of the original perturbation, and an unstable situation. As a simple analogy, the instability of a line of pullers is thus similar to the instability of a line of point charges with alternating signs where while the periodic lattice is a fixed point, any perturbation to it is unstable.}

While our  theoretical predictions agree with the numerical results obtained in Ref.~\cite{jibuti14} showing a jet instabilities for 
pullers in the absence of diffusion, we note that in contrast a recent experimental realisation of focused puller suspensions (specifically, 
the green alga  {\it Chlamydomonas}) did not display such axial clustering \cite{Martin2016}. The origin of this discrepancy  could for
instance come from the existence of a threshold in flow intensity such as the one observed in Ref.~\cite{Waisbord2016} for magnetotactic 
focusing. Indeed the {jet pearling transition} in Ref.~\cite{Waisbord2016} was obtained 
as soon as the value of the flow intensity exceeds a critical value (for a fixed external magnetic field). Another possible source of discrepancy could come from our assumption to model the swimmer as a steady puller. {\it Chlamydomonas} is  
a puller on average but in fact  oscillates between instantaneous  pusher and puller behaviours  \cite{klindt15}, potentially interfering 
with the development of an instability.

A simple extension of the situation considered in the present paper would be a configuration in which the direction of swimming is 
perpendicular to the line of swimmers. In this case, one expects pushers to be unstable while pullers would remain stable.  An other
 extension would focus  an axisymmetric situation in which a cylindrical blob of co-swimmers could be perturbed, {or multiple parallel lines of swimmers}. Such analysis would 
be closer to the experiment in Ref.~\cite{Waisbord2016} and  would be a step further towards the full modelling of instabilities of 
convected  active suspensions subject to physical taxis.    

\medskip
This work was funded in part by the European Union through a Marie Curie CIG Grant and an ERC consolidator grant to EL.

\bibliographystyle{unsrt}
\bibliography{refs_clustering}

\end{document}